\documentclass[runningheads,fleqn]{cl2emult}
\usepackage{makeidx}  
\usepackage{graphicx} 
\usepackage{subeqnar} 
\usepackage{multicol} 
\usepackage{cropmark} 
\makeindex            

\newcommand{\tr}{\mbox{tr}}
\newcommand{\dii}{\mathrm{d}}
\newcommand{\en}{{\mathcal N}}
\newcommand{\den}{\hat{\rho}}
\newcommand{\ham}{\hat{H}}

\begin{document}
\title*{Exact Conservation of Quantum Numbers in the Statistical
Description of High Energy Particle Reactions}
\toctitle{Exact Conservation of Quantum Numbers in the Statistical
Description of High Energy Particle Reactions}
\titlerunning{Exact Conservation of Quantum Numbers\ldots}
\author{
Jean Cleymans\inst{1} \and 
Antti Ker\"anen\inst{2} \and 
Esko Suhonen\inst{2}
}
\authorrunning{J. Cleymans et al.}
\institute{Department of Physics, University of Cape Town, Rondebosch
7700, South Africa
\and
Department of Physical Sciences, University of Oulu, FIN-90571 Oulu,
Finland
}
%
\maketitle

\section{Introduction}
In high energy particle collisions, the interaction
volume  is often very small and
large deviations can occur from the thermodynamic 
limit depending on the  beam energy.
In some cases it may be easy
 to produce a pion since one needs only 140 MeV
for its rest mass, however, to produce  
 an  anti-proton  one needs at least twice the rest mass, 1.88
 GeV  and  not  simply  0.94  GeV,  since 
 they can  only be produced in pairs. 
 The probability to
 produce anti-protons  therefore
 cannot   simply   follow   the  same  law  (e.g.  a  Boltzmann
 distribution) as the production of pions.  Fortunately,  statistical mechanics provides us with the tools
 to  take  into  account  constraints  like  baryon  number  or
 strangeness conservation. This will be presented in these lecture notes. 

Basically, if the system is large and hot,
the  corrections are negligible. In $Pb-Pb$ collisions 
at the CERN/SPS collider they are
negligible  because  the  
system  is  large  enough but for $p-p$ collisions at the
same  energy  these  corrections  are large and must  be  taken  into account
because the final system is too small.

The main statistical concepts are collected in Ch. 3. In chapter 3, a
statistical method for taking the exact strangeness conservation into
account is presented. This is applied in Ch. 3.1 to describe the
particle production as observed at the GSI in Ni+Ni collisions. The
exact treatment of quantum numbers is extended to include the
strangeness, baryon number and charge in Ch. 4. A comparison of
numerical results with the AGS E802 data is given in Ch. 4.1. Finally
in Ch. 5, the generalization of the method to arbitrary internal
symmetry is reviewed.

\section{Quantum Statistical Concepts in Brief}

Throughout this paper we follow the usual convention of natural units,
so the speed of light, Planck constant and Boltzmann constant have
values $c\equiv 1$, $\hbar\equiv 1$ and $k\equiv 1$, respectively.

All the physical information about a collection of particles is
contained in a density operator, $\den$. The average of an observable
$A$ in this {\em statistical ensemble} is calculated as
$\langle A\rangle=\tr(\den\hat{A})$, where $\hat{A}$ is a Hermitian operator
corresponding to the observable.
Using the density operator, one defines the entropy of the
system considered as 
$
S = -\tr(\den\ln\den).
$  
In any system in nature, the entropy is known to tend to its maximum,
so one has to find a representation of the density operator satisfying
this condition. In thermodynamical equilibrium, the average occupations of
different quantum states do not change in time, so 
$
\partial\den/\partial t = 0.
$
The density operator satisfies the equation of motion of the form
$
i \partial\den/\partial t = -[\den,\ham],
$
where $\ham$ is the Hamiltonian of the system.
Thus, the thermodynamical, stationary density operator is diagonal in the
basis formed by Hamiltonian eigenstates. 

The choice of constraints used in maximizing the entropy defines the 
type of the statistical ensemble obtained. The closed system with
fixed energy, $E$, volume, $V$, and number of particles $i$, $N_i$, is a {\em
microcanonical} ensemble. System in heat bath (the {\em ensemble
average} of energy $\langle E\rangle=\tr(\den\ham)$, $V$ and $N_i$ are
conserved) 
 corresponds to {\em canonical} ensemble. Further, if we let the
particle number $N_i$ fluctuate such that the average
$\tr(\den\hat{N_i})$ is conserved, we obtain a {\em grand canonical}
ensemble.   

Maximization of entropy using the canonical boundary conditions leads 
to the density operator of the form
$
\den = e^{-\beta \ham}/Z,
$
where $\beta$ is the inverse of temperature $T$, and $Z$ is the 
canonical {\em partition function},
\begin{equation}
Z = \tr\, e^{-\beta\ham} = 
\sum_{i} e^{-\beta E_i(N)}. 
\label{eq:z}
\end{equation}
Here $i$ labels the different quantum states in the system and $E_i(N)$
is the eigenvalue of the $N$ particle Hamiltonian.
In the last step the trace is expressed in the basis of Hamiltonian
eigenstates.
Once knowing the correct partition function, one is able to calculate
the thermodynamical quantities describing the system. For example, the
average energy is $\langle E\rangle = T^2 \partial\ln Z/\partial T$.

Choosing the grand canonical boundary conditions yields
$$
\den_G = e^{-\beta(\ham-\mu_i \hat{N}_i)}/Z_G.
$$
Here we have employed the grand canonical partition function,
\begin{equation}
Z_G(T,\{\lambda_i\},V) = \tr\, e^{-\beta(\ham - \mu_i \hat{N}_i)}
= \prod_i \sum_{N_i} \lambda_i^{N_i} Z_{N_i},
\label{eq:zg}
\end{equation}
where $\lambda_i = e^{\beta\mu_i}$ is the {\em fugacity} of the
particle species $i$ and $Z_{N_i}$ is the $N$ particle canonical
partition function  of the species $i$. Chemical potentials $\mu_i$
take care of particle number conservation in an average sense.
In this work, we are mainly interested in the mean particle numbers in
the grand canonical system, so we employ frequently the equation
\begin{equation}
\langle N_i \rangle = \lambda_i \frac{\partial \ln Z_G}{\partial
\lambda_i}.
\label{eq:N}
\end{equation}    

In properly quantized, finite volume system the partition function is
often rather difficult to compute. In this work we never know the
exact geometry of the system, so we settle for a finite volume, $V$,
sample of the infinite volume system. Thus, the summation over
discrete quantum states in partition function is changed to simple
phase space integration over continuum. The one particle canonical
partition function of particle $i$ is now given by
\begin{equation}
Z_i^1 = g_i \frac{V}{2\pi^2} \int_0^{\infty} \dii p\,  
p^2 e^{-\beta\sqrt{p^2+m_i^2}}, 
\label{eq:single}
\end{equation}
where $g_i$ is the spin degeneration factor, and $m_i$ is the mass of
the particle $i$.
Using the previous result and taking care of the correct ocupation of
quantum states, the grand canonical partition function can
be written in the form 
\begin{equation}
\ln Z_G(T,\{\lambda_i\},V) = 
\sum_i g_i \frac{V}{2\pi^2} \int_0^{\infty} \dii p\,  p^2 \ln\!
\left[ 1 + \eta_i \lambda_i e^{-\beta\sqrt{p^2+m_i^2}}\right]^{\eta_i},
\end{equation}
where $\eta$ is the statistics factor: $\eta_i = 1$ for fermions and
$\eta_i = -1$ for bosons. 
Now we can write the mean particle number as
\begin{equation}
\langle N_i \rangle = g_i \frac{V}{2\pi^2} \int_0^{\infty} \dii p\,  p^2 
\left[\lambda_i^{-1} e^{\beta\sqrt{p^2+m_i^2}} + \eta_i\right]^{-1}. 
\label{eq:num}
\end{equation}
In a rare gas (i.e. Boltzmann) limit, which is mostly applied here,
 we just put the statistics factors
$\eta_i = 0$ in particle numbers formula, or let the possible
 occupation of one particle states be only one to obtain
\begin{equation}
\ln Z_G = \sum_i \lambda_i Z_i^1.
\label{eq:singleB}
\end{equation}

In the relativistic, multispecies gas, where the conservation of
number of distinct particles is not the main interest, 
we associate the chemical potentials to conserved quantum
numbers. Given that baryon number $B$, strangeness $S$ and electric
charge $Q$ are conserved averagely in a relativistic hadron gas, the
particle numbers (\ref{eq:num}) are
\begin{equation}
\langle N_i \rangle = g_i \frac{V}{2\pi^2} \int_0^{\infty} \dii p\,  p^2 
\left[(\lambda_B^{B_i}\lambda_S^{S_i}\lambda_Q^{Q_i})^{-1} 
e^{\beta\sqrt{p^2+m_i^2}} + \eta_i\right]^{-1}, 
\label{eq:numBSQ}
\end{equation}
where $B_i$, $S_i$ and $Q_i$ are the quantum numbers of individual
particle species.
The average net quantum number, say baryon number, is a sum over
particle numbers weighted by the chosen quantum number,
\begin{equation}
\langle N_B \rangle = \sum_i B_i \langle N_i \rangle.
\label{eq:nB}
\end{equation}

\section{Exact Strangeness Conservation}

Let us consider first a  gas composed of neutral kaons and antikaons
and request that the overall strangeness be zero. The canonical (with
respect to strangeness) partition function is given by
\begin{equation}
Z_{S=0} = \prod_{i=0}^{\infty}
\left(\sum_{n_i=0}^{\infty}\frac{1}{n_i!}e^{-\beta\varepsilon_in_i}\right)
\left(\sum_{\overline{n}_i=0}^{\infty}\frac{1}{\overline{n}_i!}
e^{-\beta\varepsilon_i\overline{n}_i}\right)
\delta_{n_{K^0},n_{\overline{K}^0}},
\end{equation}
where $\varepsilon_i$ is the energy and $n_i$ the number of $K^0$'s in
the state denoted by $i$. The Kronecker delta ensures that the overall
strangeness is zero: $n_{K^0}-n_{\overline{K}^0} = \sum_{i=0}^{\infty}
(n_i-\overline{n}_i)= 0$. By including the $1/n_i!$ and
$1/\overline{n}_i!$ one gets the sums over all {\em distinct}
states. Replacing the sums over single particle levels by the
Boltzmannian momentum
integrals and using the Dirac representation
\begin{equation}
\delta(n-m) = \frac{1}{2\pi} \int_0^{2\pi} \dii\alpha e^{i(n-m)\alpha}
\end{equation} 
of delta function, one obtains the following result
\begin{eqnarray}
Z_{S=0} &=& \frac{1}{2\pi} \int_0^{2\pi} \dii\alpha 
\exp\!\left[\frac{V}{2\pi^2} \int_0^{\infty} \dii p\,  p^2
e^{-\beta\varepsilon_{K^0}+i\alpha}\right] \nonumber \\
&&\ \ \ \ \ \ \ \ \ \ \ \  
\times \exp\!\left[\frac{V}{2\pi^2} \int_0^{\infty} \dii p\,  p^2
e^{-\beta\varepsilon_{\overline{K}^0}-i\alpha}\right],
\end{eqnarray}
where $\varepsilon_{K^0}=\sqrt{p^2+m_K^2}$.
Applying the notation of single particle partition function (\ref{eq:single}) 
gives
\begin{equation}
Z_{S=0} = \frac{1}{2\pi} \int_0^{2\pi} \dii\alpha 
\exp\!\left(Z_{K^0}^1e^{i\alpha} +Z_{\overline{K}^0}^1e^{-i\alpha}\right).
\end{equation}
By expanding the exponentials in power series it is easy to perform
the $\alpha$ -integration to obtain
\begin{equation}
Z_{S=0} = \sum_{p=0}^{\infty} \frac{1}{(p!)^2}(Z_{K^0}^1)^p
(Z_{\overline{K}^0}^1)^p.  
\end{equation}
This series converges to a modified Bessel function
\begin{equation}
Z_{S=0} = I_0(2\sqrt{Z_{K^0}^1 Z_{\overline{K}^0}^1}), 
\end{equation}
which is generally defined by
\begin{equation}
I_{\nu}(x)=\sum_{l=0}^{\infty}\frac{1}{l!}\frac{1}{(l+\nu)!}\left(
\frac{x}{2}\right)^{2l+\nu}.
\end{equation}

\begin{figure}
        \begin{picture}(200,180)
                \put(0,0){
\includegraphics{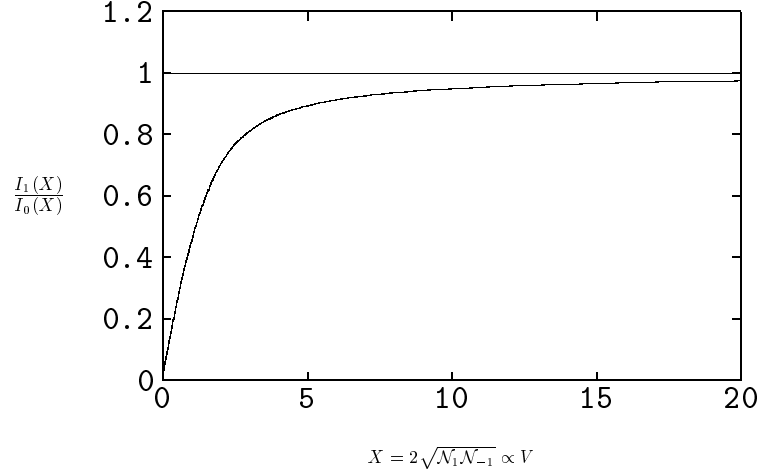}}
        \end{picture}
\caption{Nonlinear volume dependence of strange particle abundances in
canonical treatment (see eq. \ref{eq:mean1}) compared to grand
canonical case ($X\equiv 1$).} 
\label{fig:R}
\end{figure}

The generalization of the calculation to a gas containing any
particles $i$ carrying strangeness $S_i=0,\ \pm 1$ is straightforward.
For this case one obtains
\begin{equation}
Z_{S=0} = \frac{1}{2\pi} \int_0^{2\pi} \dii\alpha 
\exp\!\left(\en_1 e^{i\alpha}+\en_{-1} e^{-i\alpha}+\en_0 \right),
\label{eq:en1}
\end{equation}
where $\en_x$ stands for the sum of single particle partition
functions of particles having strangeness $S_i=x$:
\begin{equation}
\left\{
\begin{array}{lcl}
\en_1 &=& Z^1_{K^+} + Z^1_{K^0} + \ldots + Z^1_{\overline{\Lambda}}+
	   Z^1_{\overline{\Sigma}^+} + Z^1_{\overline{\Sigma}^0} +
	   \ldots \\ \\
	  
\en_{-1} &=& Z^1_{K^-} + Z^1_{\overline{K}^0} + \ldots + Z^1_{\Lambda}+
	   Z^1_{\Sigma^+} + Z^1_{\Sigma^0} +
	   \ldots \\ \\

\en_0 &=& Z^1_{\pi^+} + Z^1_{\pi^0} + Z^1_{\pi^-} + Z^1_{\eta} +
\ldots + Z^1_p + Z^1_n + \ldots
	\end{array}
\right.
\end{equation}  
In this case, where the strangeness is being treated canonically and
the baryon number and electric charge grand canonically, the
Boltzmannian one species partition function is
\begin{equation}
Z_i^1 = \lambda_B^{B_i}\lambda_Q^{Q_i}g_i 
\frac{V}{2\pi^2} \int_0^{\infty} \dii p\,  p^2
e^{\beta\sqrt{p^2+m_i^2}}.
\end{equation}
Performing the power expansion of the exponentials and the integration
in eq. (\ref{eq:en1}) we are left with
\begin{equation}
Z_{S=0} = Z_0\sum_{p=0}^{\infty}\frac{1}{p!^2}(\en_1\en_{-1})^p
= Z_0I_0(2\sqrt{\en_1\en_{-1}}),  
\label{eq:zs0}
\end{equation}
where $Z_0$ is the grand canonical partition function for particles
having strangeness zero.
To calculate the average abundancies of particles, we substitute the
fictitious strangeness fugacity and derive
\begin{equation} \label{eq:mean}
\langle N_i \rangle = \left. \lambda_i \frac{\partial \ln
Z_{S=0}}{\partial \lambda_i} \right|_{\lambda_S = 1},
\end{equation}
which gives
\begin{equation} \label{eq:mean1}
\langle N_i \rangle = Z_i^1
\left(\sqrt{\frac{\en_1}{\en_{-1}}}\right)^{S_i}
\frac{I_{S_i}(2\sqrt{\en_1\en_{-1}})}{I_0(2\sqrt{\en_1\en_{-1}})}.
\end{equation}
Each term in the sum of eq. (\ref{eq:zs0}) is the product of a
strangeness plus one and a strangeness minus one particle and one sees
the exact strangeness conservation explicitly at work. Due to the
strict conservation, the number of strange particles increases
nonlinearly with volume, which is illustrated in Fig. (\ref{fig:R}).

The method used above and the expression of the partition function
$Z_{S=0}$ indicate that to impose the strict strangeness conservation,
one projects the grand canonical partition function
$Z_G(T,\lambda_B,\lambda_Q,\lambda_S)$ onto the state with strangeness
$S$,
\begin{equation}
Z_S = \frac{1}{2\pi} \int_0^{2\pi} \dii\alpha
e^{-iS\alpha}Z_G(T,\lambda_B,\lambda_Q,\lambda_S),
\end{equation}
where the fugacity factor $\lambda_S$ has been replaced by
\begin{equation}
\lambda_S = e^{i\alpha}.
\end{equation}

The partition function for a gas containing particles with strangeness
$S_i = 0,\ \pm 1,\ \pm 2$ is given by
\begin{eqnarray}
Z_{S} &=& \frac{Z_0}{2\pi}\int_{0}^{2\pi}\dii\alpha e^{-iS\alpha}
\exp\!\left[\en_1e^{i\alpha}+\en_{-1}e^{-i\alpha}\right. \\
&& \left. + \en_{\Xi^0}(\lambda_Be^{-i2\alpha}+\lambda_B^{-1}e^{i2\alpha})
+ \en_{\Xi^-}(\lambda_B\lambda_Q^{-1}e^{-i2\alpha}+\lambda_B^{-1}
\lambda_Q e^{i2\alpha})\right], \nonumber
\end{eqnarray}
where the sums $\en_{\Xi^0}$ and $\en_{\Xi^-}$ include also the heavier
resonances carrying the same quantum numbers, as the $\en_{\pm 1}$ do.
Using the generating function for modified Bessel functions $I_{\nu}$
defined by
\begin{equation}
\exp\!\left\{\frac{\rho}{2}\left(t+\frac{1}{t}\right)\right\} 
= \sum_{\nu=-\infty}^{\infty}I_{\nu}(\rho)t^{\nu}
\end{equation}
and expanding again the exponentials  
in power series we obtain
\begin{eqnarray}
Z_{S} &=& \frac{Z_0}{2\pi}\int_{0}^{2\pi}\dii\alpha e^{-iS\alpha}
\sum_{m=-\infty}^{\infty}I_m(2\en_{\Xi^0})\lambda_B^m e^{-i2m\alpha}
\\
&&\times \sum_{n=-\infty}^{\infty}I_n(2\en_{\Xi^-})\lambda_B^n\lambda_Q^{-n}
e^{-i2n\alpha} 
 \sum_{p=0}^{\infty}\frac{1}{p!}\en_1^pe^{ip\alpha}
\sum_{q=0}^{\infty}\frac{1}{q!}\en_{-1}^qe^{-iq\alpha}. \nonumber
\end{eqnarray}
Carrying out the integration and rearranging the summations the result
can be expressed in terms of $I_{\nu}$ -functions:
\begin{eqnarray}
Z_{S} &=& Z_0\sum_{m=-\infty}^{\infty}I_m(2\en_{\Xi^0})\lambda_B^m
\sum_{n=-\infty}^{\infty}I_n(2\en_{\Xi^-})\lambda_B^n\lambda_Q^{-n}
\nonumber \\
&& \times \sum_{p=0}^{\infty}\frac{1}{p!}\frac{1}{(-S+p-2m-2n)!}
\nonumber \\
&&\times \left(\sqrt{\en_1\en_{-1}}\right)^{-S+2p-2m-2n}
\left(\frac{\en_1}{\en_{-1}}\right)^{\frac{S}{2}+m+n} \nonumber \\ 
&=& Z_0\sum_{m=-\infty}^{\infty}I_m(2\en_{\Xi^0})\lambda_B^m
\sum_{n=-\infty}^{\infty}I_n(2\en_{\Xi^-})\lambda_B^n\lambda_Q^{-n}
\nonumber \\
&&\times \left(\sqrt{\frac{\en_1}{\en_{-1}}}\right)^{S+2m+2n}I_{S+2m+2n}
(2\sqrt{\en_1\en_{-1}}).
\end{eqnarray}
Using the same techiques the result can be generalized to the case,
where the $\Omega$ -like hadrons ($S_i=\pm 3$) are included as well:
\begin{eqnarray}
Z_{S} &=& Z_0\sum_{m=-\infty}^{\infty}I_m(2\en_{\Xi^0})\lambda_B^m
\sum_{n=-\infty}^{\infty}I_n(2\en_{\Xi^-})\lambda_B^n\lambda_Q^{-n}
\nonumber \\
&&\times \sum_{l=-\infty}^{\infty}I_l(2\en_{\Omega^-})\lambda_B^l
\lambda_Q^{-l} 
\nonumber \\
&&\times\left(\sqrt{\frac{\en_1}{\en_{-1}}}\right)^{S+3l+2m+2n}
I_{S+2m+2n+3l}
(2\sqrt{\en_1\en_{-1}}).
\end{eqnarray}
The mean number of hadrons $i$ with strangeness $S_i$ in the system
becomes
\begin{equation}
\langle N_i \rangle = Z_i^1 \frac{Z_{S-S_i}}{Z_S}
\left(\sqrt{\frac{\en_1}{\en_{-1}}}\right)^{-S_i}.
\end{equation}
The modified Bessel functions decrease quickly with increasing
absolute value of their indecies, so the numerical evaluation of mean
particle numbers is not cumbersome.

\subsection{Application of $Z_S$}

In a recent paper \cite{cleymans3319} we have analysed the particle
production in GSI SIS Ni+Ni experiments. We addressed especially the
abundance of kaons who can not be described by hadronic gas model in
its standard form. Although the size of the Ni system is relatively
large the corrections due to exact strangeness conservation turned out
to be crucial at low temperatures, $T < 100$ MeV, involved.
At these temperatures the width of resonances had to be taken into
account. A summary of our results for particle ratios is presented in
table 1.

\begin{table}

\caption{Particle ratios given by present model compared to 
experimental results. The best fit value,
$\mu_B = 0.72$ GeV, for the baryon chemical potential is used.}

\begin{center}
\begin{tabular}{l|llll|lll} \hline\hline \\
Ratio   & Model    &    &          &    & Data          & \\
        & $R=4.2$ fm &  & $R=3$ fm &    &               & \\
        & $T=65$ MeV&$75$ MeV&$T=65$ MeV&$75$ MeV& ratio& 
        ref. \\ \hline 
${\rm K}^+/{\rm K}^-$  &$25.7$&$22.4$&$23.9$&$21.1$& $21\pm 9$ &\cite{KaoS,oeschler,FOPI}\\ 
${\rm K}^+/\pi^+$  &$0.0071$&$0.0339$&$0.0027$&$0.0132$& $0.0074\pm 0.0021$&\cite{oeschler,best,muentz}\\
$\phi/{\rm K}^-$   &$0.103$&$0.082$&$0.276$&$0.212$& $0.1$     &\cite{herrmann} \\
$\pi^+/\pi^-$      &$0.893$&$0.895$&$0.894$&$0.898$& $0.89$    &\cite{pelte} \\
$\eta/\pi^0$    &$0.008$&$0.015$&$0.008$&$0.015$& $0.037\pm 0.002$
        &\cite{TAPS} \\
$\pi^+/{\rm   p}$  &$0.225$&$0.247$&$0.224$&$0.246$&  $0.195\pm
0.020$&\cite{muentz,pelte}\\
$\pi^0/{\rm B}$    &$0.104$&$0.108$&$0.104$&$0.107$& $0.125\pm 0.007$
        &\cite{TAPS} \\
${\rm d}/{\rm p}$  &$0.129$&$0.188$&$0.129$&$0.188$& $0.26$    &\cite{muentz} 
\\ \hline\hline 
\end{tabular}
\end{center}

\end{table}

The measured hadronic ratios with corresponding errorbars are
described as bands in the $(T,\mu_B)$ plane as shown in Fig. 1. The
intervals of temperature and of chemical potential
\begin{eqnarray}
T &=& 70\pm 10 {\rm MeV} \nonumber \\
\mu_B &=& 720\pm 30 {\rm MeV} \nonumber
\end{eqnarray}
give a good fit to the data. The freeze-out radius of $R\simeq 4$ fm
was extracted from the volume dependence of the ratios $K^+/\pi^+$ and
$\phi/K^-$.

\begin{figure}
\begin{picture}(200,210)
\put(0,0){\includegraphics{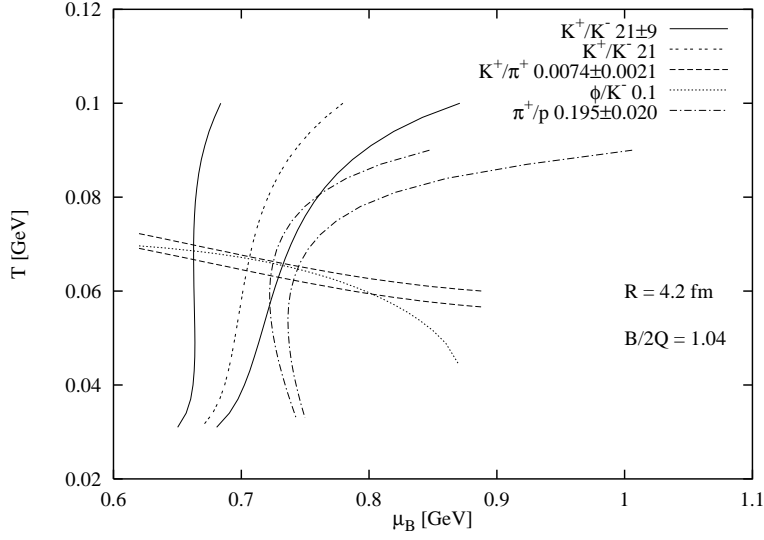}}
        \end{picture}
\caption{Curves  in  the
$(\mu_B,T)$ plane corresponding to the  hadronic ratios
indicated.
The interaction volume corresponds to the radius of 4.2 fm, and the 
isospin asymmetry
is $B/2Q = 1.04$.}
\label{fig1}
\end{figure}

\section{Exact Baryon, Charge and Strangeness Conservation}
In the case of three exactly conserved, additive quantum numbers we
start from the single particle partition function of particle $i$,
\begin{equation}
Z_i = \sum_j g_j \frac{V}{2\pi^2} \int_0^{\infty} \dii p\,  
p^2 e^{-\beta\varepsilon_j} \delta_{B_j,B_i}\delta_{Q_j,Q_i}\delta_{S_j,S_i}. 
\end{equation}
Making use of the integral representation for $\delta$ -functions and
the overall conservation constraints
\begin{equation}
\sum B_i = B,\ \ \ \sum Q_i=Q,\ \ \ \sum S_i=S,
\end{equation}
the resulting intergral corresponds to a projection of the grand
canonical partition function onto the desired values of $B$, $Q$ and
$S$:
\begin{eqnarray}
Z_{B,Q,S} &=& \frac{1}{(2\pi)^3}  \int_0^{2\pi}\dii\psi e^{-iB\psi}
				\int_0^{2\pi}\dii\phi e^{-iQ\phi}
	\int_0^{2\pi}\dii\alpha e^{-iS\alpha} \nonumber \\
&& \times		Z_G(T,\lambda_B,\lambda_Q,\lambda_S).
\label{eq:int3}
\end{eqnarray}
Here the fugacity factors have been replaced by
\begin{equation}
\lambda_B=e^{i\psi},\ \ \ \lambda_Q=e^{i\phi},\ \ \
\lambda_S=e^{i\alpha}.
\end{equation}
As the contributions always come pairwise for particles and
antiparticles, the fugacity factors will give rise to the cosine of
the corresponding angle. It is useful to group the particles according
to their quantum numbers. Leaving out charm, bottom and $S_i \geq 2$
particles we are left with ten categories which will be labeled by
their particle content. For instance, $\en_{K^0}$ stands for the sum of one
particle partition functions of neutral strange and antistrange mesons
($K^0,\ \overline{K}^0,\ {K^*}^0,\ \overline{K^*}^0,\ \ldots$),
$\en_{K^c}$ stands for charged strange mesons ($K^+,\ K^-,\ {K^*}^+,\
{K^*}^-,\ \ldots$), and $\en_{\Lambda}$ stands for all neutral hyperons
and antihyperons. With these notations the partition function can be
rewritten in the form
\begin{eqnarray}
Z_{B,Q,S} &=& \frac{1}{(2\pi)^3}  \int_0^{2\pi}\dii\psi e^{-iB\psi}
				\int_0^{2\pi}\dii\phi e^{-iQ\phi}
	\int_0^{2\pi}\dii\alpha e^{-iS\alpha} \\
&\times& \exp\!\left\{
	2\en_n \cos\psi+2\en_{\pi^c}\cos\phi+2\en_{K^0}\cos\alpha
	+ 2\en_{K^c}\cos(\phi+\alpha) 
	\right.			\nonumber \\
&&  + 2\en_p\cos(\psi+\phi) + 2\en_{\Delta^-}\cos(\psi-\phi) 
   + 2\en_{\Delta^{++}} \cos(\psi+2\phi) \nonumber \\
&& +2\en_{\Lambda}\cos(\psi-\alpha) \nonumber \\
&& \left. +2\en_{\Sigma^+}\cos(\psi+\phi-\alpha)   
   +2\en_{\Sigma^-}\cos(\psi-\phi-\alpha)\right\}. \nonumber 
\end{eqnarray}

The integration above can not be done directly due to cosine terms of
multiple angles. To circumvent this difficulty, we introduce a new
angle whenever more than one appears. For example, in the term
involving $\en_p$ we introduce an intermediate angle $\xi$ in the
following way
\begin{equation}
1 =
\int_0^{2\pi}\dii\xi\delta(\psi+\phi-\xi)=\sum_{\nu=-\infty}^{\infty}
\frac{1}{2\pi}\int_0^{\pi}\dii\xi e^{i\nu(\psi+\phi-\xi)}.
\end{equation}
The application of the integral representation of the modified Bessel function,
\begin{equation}
I_n(z) = \frac{1}{\pi}\int_0^{\pi}\dii\omega\, e^{z\cos\omega}\cos n\omega,
\end{equation}
allows one to write the partition function in the form
\begin{eqnarray}
Z_{B,Q,S}(T,V) &=& Z_0\left(\prod_{\nu=1}^{7}
\sum_{n_\nu=-\infty}^{\infty}\right) \nonumber \\
&\times&I_{-B+n_2+n_3+n_4+n_5+n_6+n_7}(2\en_n)\nonumber \\
&\times&I_{-Q+n_1+n_2-n_3+n_5-n_6+2n_7}(2\en_{\pi^c})  \\
&\times&I_{-S+n_1-n_4-n_5-n_6}(2\en_{K^0}) \nonumber \\
&\times&I_{n_1}(2\en_{K^c})I_{n_2}(2\en_{p})I_{n_3}(2\en_{\Delta^-})
\nonumber \\
&\times& I_{n_4}(2\en_{\Lambda})
I_{n_5}(2\en_{\Sigma^+})I_{n_6}(2\en_{\Sigma^-})I_{n_7}(2\en_{\Delta^{++}}).
\nonumber
\end{eqnarray}

The differentiation of eq. (\ref{eq:int3}) for particle abundances
yields the result
\begin{equation}
\langle N_i \rangle = \frac{Z_{B-B_i,Q-Q_i,S-S_i}}{Z_{B,Q,S}} Z_i^1 .
\end{equation}

The evaluation of the canonical partition function with three
simultaneously conserved quantum numbers becomes numerically very time
consuming for large values of $B$. So far, for systems with $B > 20$
we have been forced to resort to the grand canonical treatment.

\subsection{Application of $Z_{B,Q,S}$}

In order to compare our numerical results with the E802 experimental data
shown in Table \ref{tab:AGS}, we estimate the baryon number and charge of the
experimental system via geometrical considerations. Letting $R_P$ and $R_T$
be the radius of a projectile and target nucleus respectively, we assume
that the radii are directly proportional to the cubic roots of the mass
numbers A$_P$ and A$_T$ of the interacting nuclei. In the case of central
collisions the interaction region is taken to be a cylinder of radius $R_P$,
length $2\sqrt{R_T^2-R_P^2}$ plus, two remaining spherical segments at the
ends of the cylinder with $R_T$ as the radius. We find the total number of
participating nucleons (or, baryon number, $B$) and the total charge $Q$ to
be 
\begin{equation}
B=A_P+A_T\left\{ 1-\left[ 1-\left( \frac{A_P}{A_T}\right) ^{\frac 23}\right]
^{\frac 32}\right\}  \label{antti1}
\end{equation}
\begin{equation}
Q=Z_P+Z_T\left\{ 1-\left[ 1-\left( \frac{A_P}{A_T}\right) ^{\frac 23}\right]
^{\frac 32}\right\}  \label{antti2}
\end{equation}
respectively.

The comparison of our numerical results with the E802 data (Table 2)
of the $K^+/\pi^+$ and $K^-/\pi^-$ ratios as functions of the size of
the system or, equivalently, $B$ are shown in Fig. 3.
In the upper figure we observe that the theoretical curve for 
$B/2Q = 5/6$ approximates the $K^+/\pi^+$ data best. The theoretical
curves lie systematically above the data but drop closer as $B/2Q$
decreases towards the collision value. The effect of the isospin
asymmetry of the system is seen also in the $K^-/\pi^-$ data
comparison.
As the ratio $B/2Q$ approaches the collision value the theoretical
curves begin to approximate the data more closely.  

\begin{table}
\caption{Experimental results reported by the E802 collaboration. 
B and Q are calculated using equations \ref{antti1} and 
\ref{antti2}.} \vspace{6pt}
\begin{center}
\begin{tabular}{|l|c|c|c|c|c|c|}\hline\hline
Collision & $K^{+}/\pi ^{+}$ & Ref. & $K^{-}/\pi ^{-}$ & Ref. & $B$ & $Q$ \\ 
\hline
$p+\, ^4\! Be_9$ & 7.8$\pm $0.4\% & 
\cite{abb1,abb2} & 2.0$\pm $0.2\% & \cite{abb1} & 3.9 & 2.3 \\ 
$p+\, ^{13}\! Al_{27}$ & 9.9$\pm $0.5\% & 
\cite{abb2} &  &  & 5.4 & 3.1 \\ 
$p+\, ^{29}\! Cu_{64}$ & 10.8$\pm $0.6\% & 
\cite{abb2} &  &  & 6.9 & 3.7 \\ 
$p+\, ^{79}\! Au_{197}$ & 12.5$\pm $0.6\% & 
\cite{abb1,abb2} & 2.8$\pm $0.3\% & \cite{abb1} & 9.7
& 4.5 \\ 
$^{14}Si_{28}+\, ^{79}\! Au_{197}$ & 18.2$\pm $0.9\% & 
\cite{abb1} & 3.2$\pm $0.3\% & \cite{abb1} & 102.7 & 
44.0 \\ 
& 19.2$\pm $3\% & \cite{abb3} & 3.6$\pm $0.8\% & \cite
{abb3} &  &  \\ \hline\hline 
\end{tabular}
\end{center}
\label{tab:AGS}
\end{table}

\begin{figure}
\begin{picture}(200,340)
 \put(0,170){\includegraphics{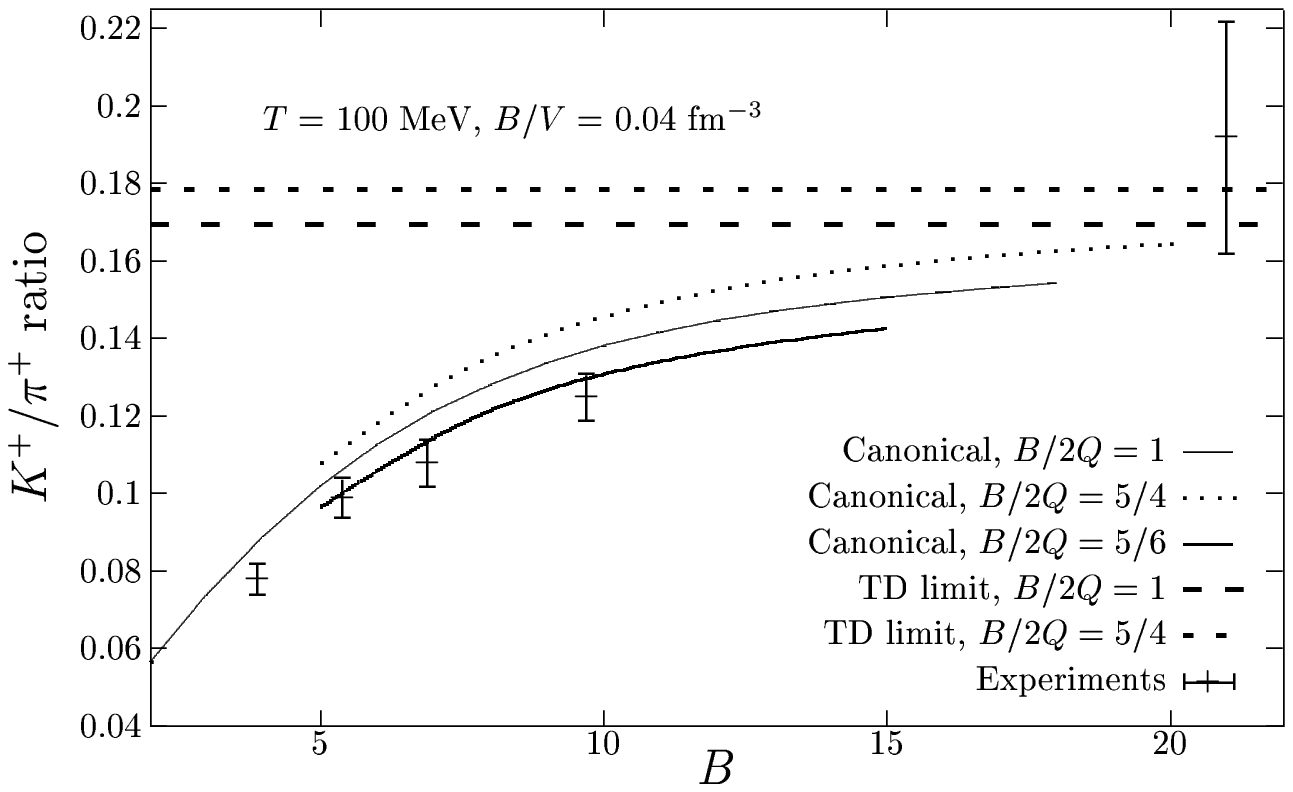}}
\put(0,0){\includegraphics{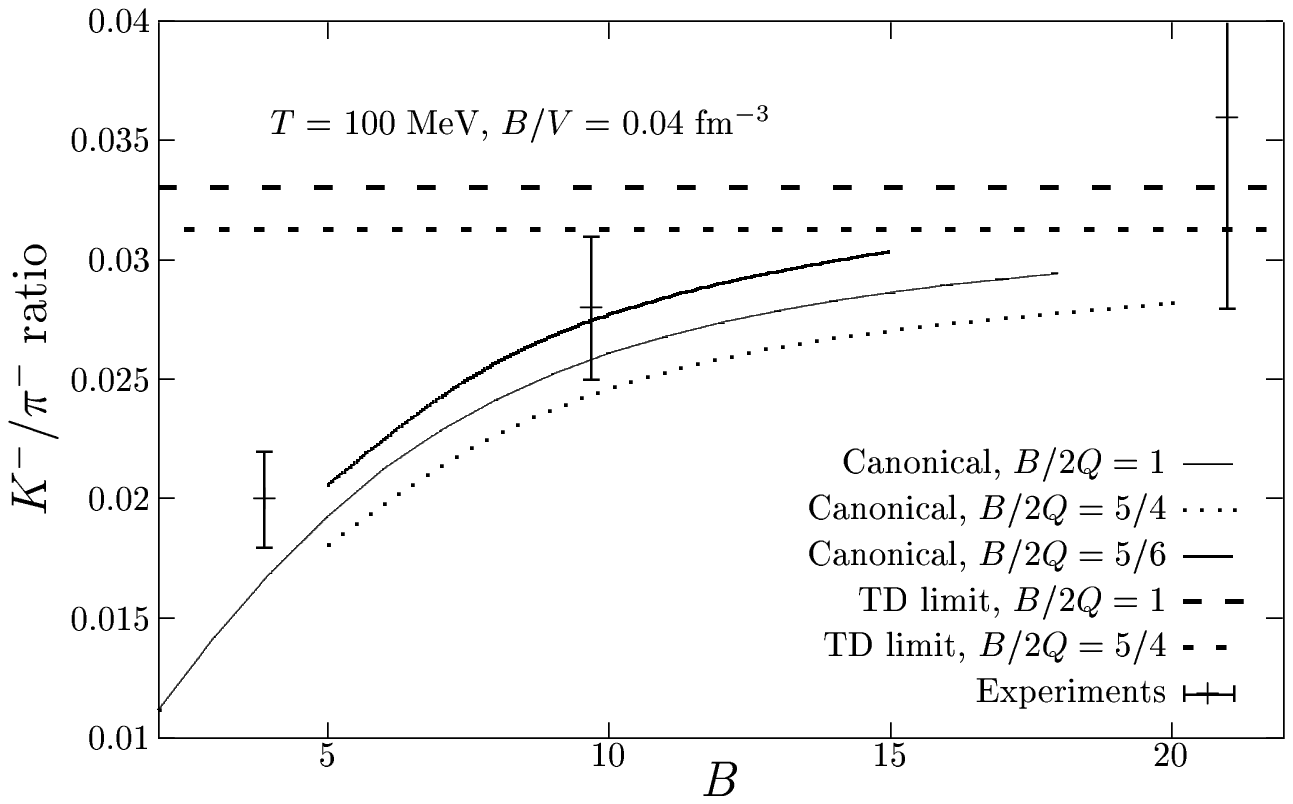}}	        
\end{picture}
\caption{
Thermal model expectations for the production ratios $K^{+}/\pi ^{+}$
and $K^{-}/\pi ^{-}$
at a temperature of 100 MeV and a baryon density of 0.04 fm$^{-3}$ compared
to experimental results from the Brookhaven AGS. The experimental
ratios from $Si-Au$ collisions ($B\sim 103$) is moved to $B=21$ for the sake
of convenience.
}
\label{fig:AGSfig8}
\end{figure}

\section{Generalization of the Projection Method}

In this section, we review the projection method generalized to
arbitrary internal symmetry of the system in addition to $U(1)$ of
strangeness and $U(1)\times U(1)\times U(1)$ of baryon number,
electric charge and strangeness. For complete derivation, see
the original texts of Turko, Redlich and Hagedorn \cite{turko}. 
The general method is suitable for
non-abelian symmetries, such as  $SU(2)$ of isospin \cite{muller274} or
angular momentum \cite{blumel637}, and $SU(3)$ of  
color \cite{elze1879} as well.

If the system is exactly symmetric under the operations of internal
symmetry group $G$, the corresponding group generators $Q_k$ have the same
eigenstates as the Hamiltonian. Thus 
\begin{equation}
[\ham, Q_k] = 0,\ k=1,\ldots,n,
\label{comm}
\end{equation}
where $n$ is the number of parameters in the group.
Let us define the {\em generating function} $\hat{Z}$ as
$
\hat{Z} = \tr[U(g)e^{-\beta\ham}],
$
where $U(g)$ is an unitary representation of the group.
With the aid of irreducible presentations of $U$, this decomposes to
\begin{equation}
\hat{Z} = \sum_{\nu} \frac{\hat{\chi}^{\nu}(g)}{d(\nu)}Z_{\nu}.
\end{equation}
Here we used a character $\hat{\chi}^{\nu}(g)$ and the dimension $d(\nu)$  
of an irreducible presentation $U^{\nu}(g)$, and the corresponding
canonical partition function $Z_{\nu}$.
Using the orthogonality of characters, 
\begin{equation}
\int\dii \mu(g) \overline{\hat{\chi}^{\nu}(g)}\hat{\chi}^{\nu'}(g) =
\delta_{\nu\nu'},
\end{equation}
we may compute the canonical partition function once we know the
generating function:
\begin{equation}
Z_{\nu} = d(\nu)\int\dii \mu(g)\overline{\hat{\chi}^{\nu}(g)} \hat{Z}.
\label{Znu}
\end{equation}
Further investigation of the generating function reveals that
\begin{eqnarray}
\hat{Z} &=& \tr \exp\!\left(-\beta \ham +
i\sum_{k=1}^{r}Q_k\gamma_k\right) \nonumber \\
&=& \prod_{j=1}^{\infty}\prod_{\rho=1}^{d(\nu)}\sum_n
\exp\!\left[n\left(-\beta E_j  +  i\sum_{k=1}^{r}
q_k^{(\rho)}\gamma_k\right) \right].
\label{hatZ}
\end{eqnarray}
In the last step, we have expressed the trace in the basis of
$n$ -particle Hamiltonian eigenstates. The $q_k^{(\rho)}$ are the
conserved charges, and the $\gamma_k$ are the variables of the Cartan
subgroup of the group $G$ of rank $r$. Eq. (\ref{hatZ}) resembles the
grand canonical partition function, and is actually obtained from it
by the Wick rotation: $\beta\mu_i \rightarrow -i\gamma_i$. 

As an example, let us choose the internal symmetry of the system
correspond to $U(1)_{q_1}\times\cdots\times U(1)_{q_r}$, where the
$q_i$ are the conserved charges. The character of $U(1)_{q_i}$ is 
$e^{iq_i\gamma_i}$, so the character of the direct product group is
$\exp(i\sum_{i=1}^r q_i\gamma_i)$. The canonical partition function
respecting the exact conservation of charges $q_i$ has now the form
\begin{eqnarray}
Z_{q_1,\ldots,q_r}(T,V) &=& 
\frac{1}{(2\pi)^r}\int_{0}^{2\pi}{\rm d}\gamma_1\cdots\int_{0}^{2\pi}
{\rm d}\gamma_r \nonumber \\
&&\times \exp\!\left[-i\sum_{i=1}^r q_i \gamma_i\right]
\hat{Z}(T,V,\gamma_1,\cdots,\gamma_r).
\end{eqnarray}
The special cases, $Z_S$ and $Z_{B,Q,S}$ for a Boltzmannian hadron
resonance  gas are considered in previous sections. 

\section{Summary}
The  particle  abundances  have  been  computed in the canonical
formalism  using  the formulation for the exact conservation of
baryon  number,  strangeness and charge in the thermal model of
particle production.
A good agreement with the experimental data of GSI Ni+Ni
collisions and of E802 collaboration in $p-A$ collisions was reported.

The  good  agreement with chemical equilibrium does not mean that
the   particle   spectra  should  follow  exactly  a  Boltzmann
distribution  since  the  momenta  of particles can be severely
affected by flow. As an example, a model with Bjorken expansion
in  the  longitudinal  direction  will  still have its particle
ratios   determined   by  Boltzmann  factors  even  though  the
longitudinal   distribution   is   nowhere   near  a  Boltzmann
distribution \cite{jaipur}.

\end{document}